\documentclass[lettersize,journal]{IEEEtran}
\usepackage{graphicx}
\usepackage{subcaption}
\usepackage{amsmath,amsfonts}
\usepackage{amssymb}
\usepackage{algorithm}
\usepackage{algpseudocode}
\usepackage{array}
\usepackage[caption=false,font=normalsize,labelfont=sf,textfont=sf]{subfig}
\usepackage{textcomp}
\usepackage{mathtools}
\usepackage{stfloats}
\usepackage{url}
\usepackage{color}
\usepackage{verbatim}
\usepackage[acronym,toc]{glossaries}
\usepackage{lipsum}
\usepackage{balance}
\usepackage{cite}
\usepackage{booktabs}

\newtheorem{theorem}{Theorem}

\newcommand{\changed}[1]{{#1}}

\newacronym{AoI}{AoI}{age of information}
\newacronym{AAoI}{AAoI}{average AoI}
\newacronym{PAoI}{PAoI}{peak AoI}
\newacronym{OFDMA}{OFDMA}{orthogonal frequency division multiple access}
\newacronym{AIRA}{AIRA}{age independent random access}
\newacronym{ADRA}{ADRA}{age dependent random access}
\newacronym{ADRA-MR}{ADRA-MR}{age dependent random access in multiple relays}
\newacronym{AIRA-MR}{\textsc{AIRA-MR}}{age independent random access in multiple relays}
\newacronym{ADRF}{ADRF}{age dependent relay forwarding}
\newacronym{CAP}{CAP}{channel access probability}
\newacronym{CSI}{CSI}{channel state information}
\newacronym{DTMC}{DTMC}{discrete time Markov chain}
\newacronym{LEO}{LEO}{low earth orbit}
\newacronym{IoT}{IoT}{Internet of Things}
\newacronym{MTC}{MTC}{machine type communications}
\newacronym{RA}{RA}{random access}
\newacronym{SIC}{SIC}{successive interference cancellation}
\newacronym{SA}{SA}{slotted ALOHA}
\newacronym{CSMA}{CSMA}{carrier sensing multiple access}
\newacronym{MIMO}{MIMO}{multiple input multiple Output}
\newacronym{NOMA}{NOMA}{non-orthogonal multiple access}
\newacronym{OMA}{OMA}{orthogonal multiple access}
\newacronym{ARQ}{ARQ}{automatic repeat request}
\newacronym{SINR}{SINR}{signal to interference plus noise ratio}
\newacronym{LPWAN}{LPWAN}{low power wide area network}
\newacronym{MR}{MR}{multiple-relay}
\newacronym{AD}{AD}{age dependent}
\newacronym{MAM}{MAM}{max-age matching}
\newacronym{RMD}{RMD}{random matching diversity}
\newacronym{MAA}{MAA}{max-age algorithms}
\newacronym{IMAS}{IMAS}{iterative max-age scheduling}
\newacronym{B-IMAS}{B-IMAS}{buffered IMAS}
\newacronym{ABDR}{ABDR}{age-based delayed request}
\newacronym{B-ABDR}{B-ABDR}{buffered ABDR}
\newacronym{MRSN}{MRSN}{multirelay satellite network}
\newacronym{CDF}{CDF}{cumulative density function}
\newacronym{UAV}{UAV}{unmanned aerial vehicle}
\newacronym{RTS}{RTS}{request to send}
\newacronym{CTS}{CTS}{clear to send}

\newglossaryentry{end device}
{
  name=ED,
  description={end device},
  first={\glsentrydesc{ED} (\glsentrytext{ED})},
  plural={EDs},
  descriptionplural={end devices},
  firstplural={\glsentrydescplural{ED} (\glsentryplural{ED})}
}
\newacronym{ED}{ED}{end device}

\newglossaryentry{access point}
{
  name=AP,
  description={access point},
  first={\glsentrydesc{AP} (\glsentrytext{AP})},
  plural={APs},
  descriptionplural={access points},
  firstplural={\glsentrydescplural{AP} (\glsentryplural{AP})}
}
\newacronym{AP}{AP}{access point}

\newglossaryentry{multirelay satellite network}
{
  name=MRSN,
  description={multirelay satellite network},
  first={\glsentrydesc{MRSN} (\glsentrytext{MRSN})},
  plural={MRSNs},
  descriptionplural={multirelay satellite networks},
  firstplural={\glsentrydescplural{MRSN} (\glsentryplural{MRSN})}
}

\newglossaryentry{unmanned aerial vehicle}
{
  name=UAV,
  description={unmanned aerial vehicle},
  first={\glsentrydesc{UAV} (\glsentrytext{UAV})},
  plural={UAVs},
  descriptionplural={unmanned aerial vehicles},
  firstplural={\glsentrydescplural{UAV} (\glsentryplural{UAV})}
}

\def\BibTeX{{\rm B\kern-.05em{\sc i\kern-.025em b}\kern-.08em
    T\kern-.1667em\lower.7ex\hbox{E}\kern-.125emX}}
    
\title{Age of Information in Multi-Relay Networks with Maximum Age Scheduling}
\author{Gabriel Martins de Jesus, Felippe Moraes Pereira, Jo{\~a}o Luiz Rebelatto,\\ Richard Demo Souza, Onel Alcaraz L{\'o}pez
\thanks{G. M. de Jesus and O. A. L{\'o}pez are with the Centre for Wireless Communications (CWC), University of Oulu, 90014 Oulu, Finland. (e-mail: gabriel.martinsdejesus@oulu.fi, onel.alcarazlopez@oulu.fi).}
\thanks{F. M. Pereira is with CPGEI, Federal University of Technology - Paran{\'a}, Curitiba 80230-901, Brazil (e-mail: felippepereira@alunos.utfpr.edu.br).}
\thanks{J. L. Rebelatto is with CPGEI and DAELN, Federal University of Technology - Paran{\'a}, Curitiba 80230-901, Brazil (e-mail: jlrebelatto@utfpr.edu.br).}
\thanks{R. D. Souza is with the Electrical and Electronics Engineering Department, Federal University of Santa Catarina (UFSC), Florian{\'o}polis 88040-900, Brazil (e-mail: richard.demo@ufsc.br).}
\thanks{This research has been supported by the Research Council of Finland (former Academy of Finland) Grant 346208 (6G Flagship Programme), CNPq (403124/2023-9, 305021/2021-4) and RNP/MCTI Brasil 6G project (01245.020548/2021-07).}}

\begin{document}
\maketitle

\begin{abstract}
We propose and evaluate age of information (AoI)-aware multiple access mechanisms for the Internet of Things (IoT) in multi-relay two-hop networks. The network considered comprises end devices (EDs) communicating with a set of relays in ALOHA fashion, with new information packets to be potentially transmitted every time slot. The relays, in turn, forward the collected packets to an access point (AP), the final destination of the information generated by the EDs. More specifically, in this work we investigate the performance of four age-aware algorithms that prioritize older packets to be transmitted, namely max-age matching (MAM), iterative max-age scheduling (IMAS), age-based delayed request (ABDR), and buffered ABDR (B-ABDR). The former two algorithms are adapted into the multi-relay setup from previous research, and achieve satisfactory \changed{average AoI and average peak AoI} performance, at the expense of a significant amount of information exchange between the relays and the AP. The latter two algorithms are newly proposed to let relays decide which one(s) will transmit in a given time slot, requiring less signaling than the former algorithms. We provide an analytical formulation for the AoI lower bound performance, compare the performance of all algorithms in this set-up, and show that they approach the lower bound. The latter holds especially true for B-ABDR, which approaches the lower bound the most closely, tilting the scale in its favor, as it also requires far less signaling than MAM and IMAS.
\end{abstract}


\section{Introduction}

\IEEEPARstart{A}{} significant challenge in the \gls{IoT} landscape is to provide reliable and ubiquitous connectivity to low-complexity devices in remote areas~\cite{Centenaro:ICST:2021, Nguyen:JIOT:2022}. The use of two-hop multi-relay networks is an attractive solution due to their ability to provide seamless connectivity, even in areas where the \gls{AP} is far from the \glspl{ED}. \changed{Moreover, the use of relays may reduce operational costs, compared to upgrading the \glspl{ED} or deploying \glspl{AP} closer to them~\cite{Munari:ITC:2021}.} In these networks, the communication process is usually divided into two phases: In {phase-1}, \glspl{ED} connect to relays, which in turn communicate with the corresponding \gls{AP} during {phase-2}. Although time and code-based division multiple access techniques (TDMA and CDMA, respectively) can be used to connect the \glspl{ED} to the relays, \changed{random access protocols like ALOHA are} more practical for setups with a potentially large number of \glspl{ED}. This is because these devices typically have limited power and computational capacities, and the cost of coordinating them would quickly overcome the potential benefits. Indeed, such coordination reduces the number of lost packets due to collisions or miss detections, but it negatively impacts the timely delivery of packets and/or significantly increases the energy consumption of the \glspl{ED}.

The coordination of \glspl{ED} may be infeasible or disadvantageous, but {coordination} among relays is a more practical task, \changed{and it may lead to performance improvements. The quantity of relays is much smaller than that of \glspl{ED}, and it is more reasonable that the \gls{AP} can exchange information with the relays before packet transmission to coordinate them. Moreover, depending on the setup, it may be possible for relays to communicate among themselves, coordinating independently from the \gls{AP}. The specific protocol implemented in phase-2 must be selected to comply with the system requirements, but the system performance is inevitably limited by how many packets are captured in phase-1. However, as relays are capable of complex operations, the algorithms for forwarding the received packets can be developed to approach the performance bound set by phase-1.}

When developing these algorithms, the performance metric to be optimized must be carefully selected. For example,~\cite{Munari:ITC:2021} considered throughput, a useful metric for systems that demand a high flux of information. Meanwhile, for applications with timely demands, quantifying the \textit{freshness} of the information at the \gls{AP} becomes of paramount importance. Unlike traditional latency metrics, which focus on the delay between transmission and reception, the so-called \gls{AoI} \cite{Kaul:ISIT:2017,Yates:JSAC:2021} measures the age of the most recent data update from the perspective of the \gls{AP}. This is particularly important in applications such as real-time monitoring, industrial automation, and \gls{IoT}, where the freshness of information is critical for decision-making. By minimizing \gls{AoI}, systems can ensure that decisions are based on the most current data, improving reliability and performance in dynamic environments. \changed{In single-hop setups, the \gls{AoI} must be minimized with algorithms and/or protocols to be executed and/or implemented in the interface between the \glspl{ED} and the {AP}, which may require that the \glspl{ED} perform complex operations. Conversely, a multi-hop network removes this complexity from the typically inexpensive \glspl{ED}. It transfers it to the relays, which not only are more powerful but also far fewer than \glspl{ED}, even encouraging coordination with fewer drawbacks. However, the two-hop configuration also comes with limitations and its performance is generally limited by the performance achieved in phase-1, the \glspl{ED} to relays link. Moreover, feedbacks from the central server require two hops to reach the \glspl{ED}, possibly compromising the timeliness of the updates and also forbidding that {EDs} take action to control their own \gls{AoI}.}

\subsection{Related Work}
Recent works have addressed the \gls{AoI} performance of both single hop~\cite{Grybosi:JIOT:2022,Tripathi:GLOBECOM:2017,Wang:ITC:2024,Wang:ITII:2023,Moradian:ITC:2024,Chen:ICCC:2020,Yavascan:2021:IJSAC,Yu:ITWC:2023,Agarwal:ITVT:2024} and relay-assisted two-hop \gls{IoT} networks~ \cite{Chiarotti:ITC:2024,Pan:ITVT:2023,Xie:ITWC:2024,Kahraman:IOTJ:2024,deJesus:IEEEAccess:2022,deJesus:EuCNC:2023,Wang:ITVT:2024,Liu:ITNSE:2024, Cai:WCNC:2024,Yuan:IJSAC:2024,Zhou:WCNC:2024}. In \cite{Grybosi:JIOT:2022}, the authors evaluate the \gls{AoI} performance of a \gls{SIC}-aided multiple access network where the AP is capable of resolving collisions. Although the scheme from \cite{Grybosi:JIOT:2022} reduces the \gls{AoI}, it requires advanced \gls{CSI} of multiple superimposed packets to perform \gls{SIC}, which may be a prohibitive assumption in two-hop multi-relay networks.
Considering feedback between the \gls{AP} and \glspl{ED} and multiple independent channels, \cite{Tripathi:GLOBECOM:2017} addresses the scheduling between \glspl{ED} and channels using a proper matching between the instantaneous \gls{AoI} of the \glspl{ED} and their channel conditions by introducing the \gls{MAM} algorithm. \changed{\gls{MAM} is asymptotically optimal in the single-hop scenario, achieving optimal performance as $N\to\infty$ and $F=\Omega(\log N)$, where $N$ is the number of \glspl{ED} and $F$ is the number of available channels.} The authors also propose an iterative version of the algorithm \changed{that is less  computationally complex,} the near-optimal \gls{IMAS} algorithm.

The authors in \cite{Wang:ITC:2024} adopt a convex optimization strategy to minimize the \gls{AAoI} by optimizing the channel access probability of \glspl{ED}, thus reducing the \gls{AoI} by more than $50~\%$ and $20~\%$ compared to ALOHA and CSMA, respectively.
In \cite{Wang:ITII:2023}, the authors derive age-aware algorithms for systems where two classes of \glspl{ED} share the same network infrastructure. The classes reflect the activation patterns of the \glspl{ED}, either periodic or random. They propose algorithms based on Lyapunov optimization and deep reinforcement learning, with considerable performance improvement given a few devices. 

The length of the transmission frames is dynamically adapted in \cite{Moradian:ITC:2024} based on the \gls{AoI} of \glspl{ED} and the potential \gls{AoI} reduction in the current slot. Information on the frame length is broadcast to \glspl{ED} alongside an age threshold that the \glspl{ED} must comply with to start a transmission, similarly to~\cite{Chen:ICCC:2020, Yavascan:2021:IJSAC}. The proposed method successfully minimizes the \gls{AoI} increase when compared to other algorithms with fixed frame length and age threshold. The length of the packet is dynamically adapted in \cite{Yu:ITWC:2023} based on the \gls{CSI} to minimize the \gls{AoI}, improving the \gls{AAoI} even given strict power constraints.
The performance of \gls{NOMA} is evaluated in \cite{Agarwal:ITVT:2024}, and an \gls{ED} pairing scheme that schedules one or two \glspl{ED} at each time slot is proposed to minimize the \gls{AAoI}. The \glspl{ED} are selected independently of their \gls{AoI}, based solely on the transmission signal-to-noise ratio.

In two-hop setups, the case with a single source has been studied in the literature from the \gls{AoI} perspective in \cite{Chiarotti:ITC:2024,Pan:ITVT:2023,Xie:ITWC:2024}. The authors in \cite{Chiarotti:ITC:2024} consider a network consisting of a single source, a relay and a destination node. When active, the source node transmits a packet over an erasure-prone channel. If the packet is erased at the destination node, but not at the relay, the relay can transmit it in an orthogonal channel. They propose a game-theoretical representation of the system, showing that it is possible to optimize the parameters to minimize the \gls{AoI} without signaling of the nodes involved. TDMA and \gls{NOMA} are evaluated  in \cite{Pan:ITVT:2023} considering a similar setup. In the TDMA case, the source and relay alternate time slots to avoid collisions. In the \gls{NOMA} case, both the source and the relay transmit whenever possible. While {TDMA} guarantees a lower energy consumption, \gls{NOMA} offers a lower \gls{AAoI}, but the use of the relay is only beneficial when the quality of the channel is low. The authors in \cite{Xie:ITWC:2024} consider a system model that differs from~\cite{Pan:ITVT:2023,Chiarotti:ITC:2024} by having multiple relays and no direct link between \glspl{ED} and the destination node. The relays act cooperatively at each time slot, and the first $k$ relays that reported receiving the packet are selected to forward it. Either selection combining or maximal-ratio combining is performed at the destination. Analytical formulations are used to optimize $k$, as well as the packet lengths in the first and second hops. 

Conversely, the \gls{AoI} in multi-source setups with two hops has been studied in \cite{Kahraman:IOTJ:2024,deJesus:IEEEAccess:2022,deJesus:EuCNC:2023,Zhou:WCNC:2024,Wang:ITVT:2024,Liu:ITNSE:2024, Cai:WCNC:2024,Yuan:IJSAC:2024}. 
Specifically, network coding is studied in \cite{Kahraman:IOTJ:2024}, where the authors present analytical expressions for the \gls{AAoI} of \glspl{ED} and show significant performance gains compared to uncoded networks. Building on~\cite{Chen:ICCC:2020,Yu:TWC:2021}, the authors in \cite{deJesus:IEEEAccess:2022,deJesus:EuCNC:2023} extend the setup from~\cite{Munari:3BCCN:2019,Munari:ITC:2021} and propose \gls{AoI}-based multiple access schemes that significantly reduce the \gls{AAoI} compared to a pure slotted-ALOHA scheme, but requiring a feedback link from the \gls{AP} all the way back to the \glspl{ED}, which may be prohibitive in practice.

A typical two-hop \gls{IoT} setup is that where \glspl{UAV} work as relays for \glspl{ED}. The authors in \cite{Wang:ITVT:2024} propose \gls{UAV} flying patterns with the aim of minimize the \gls{AoI} of edge users, while achieving a performance closer to optimal compared to previous methods. In \cite{Liu:ITNSE:2024}, \glspl{UAV} relay \glspl{ED} packets to \gls{LEO} satellites, while also supplying power to \glspl{ED}. The positions and trajectory of the \glspl{UAV} are determined after solving several optimization subproblems, and the \gls{AAoI} is nearly optimal. In \cite{Cai:WCNC:2024}, a single satellite works as a relay for a set of \glspl{ED} and a scheduling algorithm is proposed to minimize the \gls{PAoI}, outperforming the benchmarks considered. The authors in \cite{Yuan:IJSAC:2024} evaluate the \gls{AAoI} in an earth-moon communication setup with satellite relays. They propose a satellite constellation scheme, with the system's parameters, such as satellite height, optimized to minimize the \gls{AAoI} per device.

\subsection{Contribution}
In this paper, we investigate max-age algorithms in two-hop multi-relay networks. First, we extend the \gls{MAM} and \gls{IMAS} algorithms from \cite{Tripathi:GLOBECOM:2017} to the two-hop setup.  Then, we propose two distributed methods that achieve \gls{AoI} results comparable to those of the former algorithms with much less information exchange before transmission. More specifically, the contributions of this paper can be listed as follows:
\begin{enumerate}
    \item We develop an analytical lower-bound for the \gls{AAoI} and \gls{PAoI} of our setup, which are validated by simulation results. This formulation can be used to properly select system parameters, such as the activation probability, number of \glspl{ED}, relays, and frequency channels, for a given erasure rate and a desired performance level.  
    \item We extend the single-hop \gls{MAM} and \gls{IMAS} schemes from~\cite{Tripathi:GLOBECOM:2017} to the two-hop setup by shifting the packet transmission selection to the relays instead of the \glspl{ED}. In case of multiple relays capturing the same packet, only one forwards it, avoiding redundancy. 
    \item In order to reduce the signaling demanded by MAM/IMAS, we propose a scheme called \gls{ABDR}, where the relays themselves resolve the forwarding of packets instead of relying on a centralized scheduling done solely by the \gls{AP}. 
    \item We enhance the \gls{ABDR} scheme by enabling relays to retain packets that were not transmitted, allowing their transmission in subsequent time slots at more opportune channel conditions. The so-called \gls{B-ABDR} scheme achieves performance significantly close to the lower bound, while requiring the same amount of information exchange as \gls{ABDR}.
\end{enumerate}

The remainder of this paper is organized as follows. In Section~\ref{sec:system_model}, we present our system model and the main performance metrics considered in this work. Section \ref{sec:LowerBound} presents the analytical formulation for the \gls{AoI} lower bound in the two-hop setup. In Section \ref{sec:MAA}, we discuss the max-age algorithms and their extension to our setup, while Section \ref{sec:ABDR} introduces new distributed scheduling schemes. In Section \ref{sec:results}, we present some simulation results and compare the algorithms discussed with the analytical lower bound. Lastly, Section \ref{sec:conclusions} concludes the paper. 
In Table \ref{tab:list_of_symbols}, we present a list of the symbols used in this paper and their respective meaning.

\begin{table}[!t]
    \centering
    \caption{List of symbols used in the paper.}
    \begin{tabular}{c l c}
    \hline
      Symbol               &  Meaning  & Default value in Sec. \ref{sec:results}\\
    \hline
      $N$                  & Number of \glspl{ED} & $30$\\
      $p$                  & Activation probability & $0.1$\\
      $F$                  & Number of channels & $2$\\
      $K$                  & Number of relays & $5$\\
      $\varepsilon_1$      & Erasure rate in phase-1 & $0.1$\\
      $\varepsilon_2$      & Erasure rate in phase-2 & $0.1$\\
      $\Delta_i(t)$        & \gls{AoI} of the $i$-th \gls{ED} & -\\
      $\bar{\Delta}$       & \gls{AAoI} of the network & -\\
      $\bar{\Delta}^{(p)}$ & \gls{PAoI} of the network & -\\
      $\pi(a)$             & Stationary dist. of the \gls{AoI} of \glspl{ED} & -\\
      $Q$                  & Prob. of success of transmission & -\\      
    \hline
    \end{tabular}
    \label{tab:list_of_symbols}
\end{table}

\section{System Model}\label{sec:system_model}
We consider a network composed of $N$ \glspl{ED} that transmit their data packets to a common \gls{AP}. Time is divided into transmission opportunities, referred to as time slots, where each one is sufficient for the generation, transmission, reception, and processing of a packet. Following a generate-at-will policy~\cite{Chen:ICCC:2020}, the \glspl{ED} monitor their environment and generate packets with probability $p$, assumed to be equal among all \glspl{ED}. \changed{We consider low-cost \glspl{ED} with limited computation and power capabilities. For this reason, the channel access is slotted ALOHA, where \glspl{ED}} transmit the packet in one out of the $F$ shared frequency channels, which is selected randomly and uniformly. After being transmitted by an \gls{ED}, the packet is discarded (there is no retransmission).

As in \cite{Munari:ITC:2021,deJesus:IEEEAccess:2022,deJesus:EuCNC:2023}, we assume that there is no direct link between \glspl{ED} and \gls{AP}, such that the connection is realized with the aid of a set of $K$ relays, which receive the \glspl{ED}' packets and forward them to the \gls{AP}. This setup is representative, e.g., of a \gls{LEO} satellite network~\cite{Qu:Access:2017} and of star-of-stars topology networks, such as low power wide area networks~\cite{Raza:CST:2017}. 

We denote by $\mathcal{R}$ the list of indexes of users whose packets were captured by the relays in a given time slot. When a packet is received by a relay, it can be forwarded to the \gls{AP} according to the policy in use. The trivial decision to always forward packets results in several collisions at the \gls{AP}, encouraging the development of policies that limit the number of forwarded packets~\cite{deJesus:IEEEAccess:2022,deJesus:EuCNC:2023}. Here, we consider that the \glspl{ED} are low-complexity devices, while relays are capable of sophisticated exchanges with the \gls{AP}. Thus, while there is no feedback link from the relays to \glspl{ED}, the relays and \gls{AP} can exchange information, such as pilot sequences as well as instantaneous \gls{AoI} of users. After the exchange phase with the \gls{AP}, a list of the channels that the relays will use is obtained, denoted by $\mathcal{F}^\star$.

\begin{figure}[!t]
\centering
\includegraphics[width=0.5\textwidth]{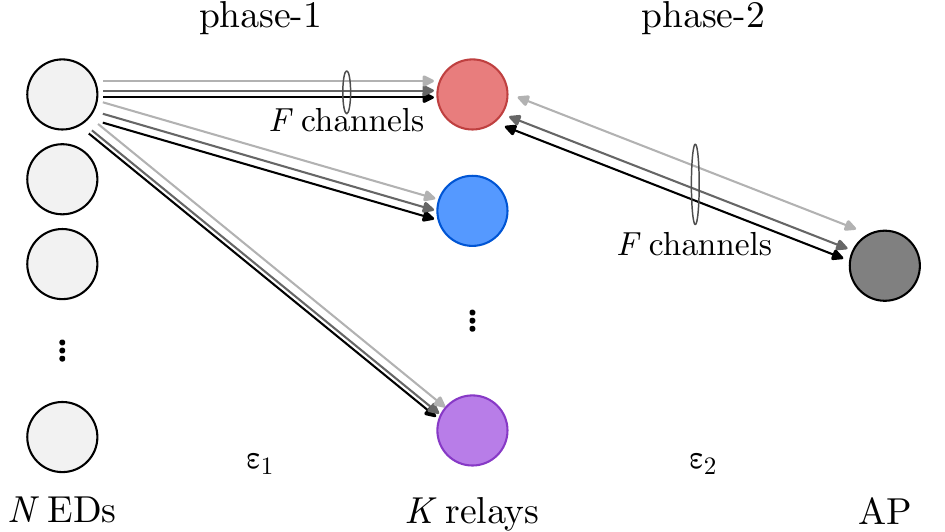}%
\caption{Illustration of the system model. Arrows indicate the direction of the connection, e.g., \glspl{ED} only transmit to relays, while relays can transmit/receive information to/from the \gls{AP}.}
\label{fig:SystemModel}
\end{figure}

The \glspl{ED} share $F$ channels to the relays, each characterized by the on-off fading model~\cite{Perron:ISTS:2003}, such that a packet is erased with probability $\varepsilon_1$. Moreover, the relays have access to $F$ orthogonal channels to the \gls{AP}, also characterized by the on-off fading model but with erasure probability $\varepsilon_2$. The on-off fading model is convenient as it is mathematically tractable and captures channel impairments such as fading and obstructions from physical obstacles. This model has been used extensively in the \gls{IoT} literature, e.g., \cite{Munari:ITC:2021} and references therein. The connectivity between the relays and \gls{AP} can be represented by a connectivity matrix $\mathbf{H}$, where each element $h_{f,k}$ represents the connectivity of the $k$-th relay to the \gls{AP} in the $f$-th channel, and is valued either 0 or 1. At any reception point (relays or \gls{AP}), an arriving packet is not received either due to being erased or due to a collision, i.e., when more than one packet is transmitted and not erased simultaneously. At the end of each time slot, the \gls{AP} broadcasts to relays a list of all packets successfully delivered to the \gls{AP} in that time slot. Fig.~\ref{fig:SystemModel} presents an illustration of the system model, where we only explicitly show the connections of one \gls{ED} to all relays, and the connection of one relay to the \gls{AP}.

\subsection{Age of Information} 

In some remote sensing applications, the \textit{freshest} information, meaning the one generated most recently, is generally desired. By using the \gls{AoI}~\cite{Kaul:INFOCON:2012}, the freshness of information is quantified, and the quality of a system can be objectively evaluated with this quantity. At the $t$-th time slot, let $r_i(t)$ be the time slot where the last packet received by the \gls{AP} from the $i$-th \gls{ED} was generated. Then, the instantaneous \gls{AoI} of this device is defined as a random process given by $\Delta_i(t) = t - r_i(t)$~\cite{Kaul:ISIT:2017}. In a discrete-time system, an \gls{ED}'s \gls{AoI} increases by one at each time slot without new packets arriving at \gls{AP}. However, when a packet is collected, the \gls{AoI} is reset to the age of such packet. The age of the packet, similarly, is given by the difference between the current time slot and the time slot the packet was generated. In a system with no retransmissions and no buffering, this difference is 0, such that, in the next time slot, the \gls{ED}'s \gls{AoI} is 1. 

We denote by $\{\Delta_i(t)\}_{i=1}^N = \{\Delta_1(t), \Delta_2(t), \dots, \Delta_N(t)\}$ the set of instantaneous \gls{AoI} of all the $N$ \glspl{ED} at time slot $t$.
The \gls{AAoI} of the $i$-th \gls{ED} is obtained as~\cite{Chen:ICCC:2020}
\begin{equation}\label{eq:avg_aoi_lim}
    \bar{\Delta}_i =  \lim_{T \rightarrow \infty} \frac{1}{T}\sum_{t=1}^{T}\Delta_i(t).
\end{equation}
Then, the network \gls{AAoI} is determined by averaging the \gls{AAoI} of all \glspl{ED} as
$\bar{\Delta} = \frac{1}{N}\sum_{i=1}^N \bar{\Delta}_i$. Since all \glspl{ED} are assumed to experience the same system statistics, we have that $\bar{\Delta} = \bar{\Delta}_i$.
Besides \gls{AAoI}, one can rely on \gls{PAoI} to evaluate the performance of the system with respect to the maximum \gls{AoI} an \gls{ED} achieves on average, designing the network such that a threshold \gls{AoI} is achieved with a given probability \cite{Costa:ITIT:2016}. The \gls{PAoI} measures the average value of an \gls{ED}'s \gls{AoI} immediately before it has its packet delivered to the \gls{AP}. Let $\mathcal{T}_i$ denote the set of time slots where the $i$-th \gls{ED} had its packet successfully delivered to the \gls{AP}. Then, the $i$-th \gls{ED} \gls{PAoI} is given by
\begin{equation}
    \bar{\Delta}^{(p)}_i = \frac{1}{|\mathcal{T}_i|}\sum_{t\in\mathcal{T}_i}\Delta_i(t),
\end{equation}
where $|\mathcal{T}_i|$ is the cardinality of $\mathcal{T}_i$. Meanwhile, the network \gls{PAoI} is given by the average of the \gls{PAoI} of all \glspl{ED} as $
    \bar{\Delta}^{(p)} = \frac{1}{N}\sum_{i=1}^N \bar{\Delta}^{(p)}_i$. As before, $\bar{\Delta}^{(p)} = \bar{\Delta}^{(p)}_i.$

The evaluation of \gls{AAoI} and \gls{PAoI} is not often straightforward. Interestingly, the authors in~\cite{Chen:ICCC:2020} model the evolution of the \gls{AoI} of the \glspl{ED} as a \gls{DTMC} and we rely on their formulation to support part of our analysis. When an \gls{ED} transmits a packet and it has current \gls{AoI} $a$, $q(a)$ denotes the probability of a packet to be successfully delivered to the \gls{AP}. For any \gls{ED}, ommiting the ED index and conditioning $\Delta(t) = a$, the \gls{AoI} evolves as \begin{equation}
    \Pr\left\{\Delta(t+1) = a+1\right\} = 1-p\,q(a),\end{equation}
    and
    \begin{equation}
    \,\Pr\left\{\Delta(t+1)=1\right\} = p\,q(a).
\end{equation}
This \gls{DTMC} is irreducible, and there exists a stationary distribution $\mathbf{\pi}(a)$, with each element $\pi(a)$ standing for the probability of the \gls{ED} to have $\Delta(t) = a$ for any $t$.

In systems where the \gls{AoI} is not used to determine the scheduling, $q(a) = Q$ is constant and independent of $\Delta(t)$, and each element of $\mathbf{\pi}(a)$ is given by \cite{Chen:ICCC:2020}
\begin{equation}\label{eq:pdf_aoi_eds}
    \pi(a) = p\,Q(1-p\,Q)^{a-1}.
\end{equation}
The \gls{AAoI} is then calculated as
\begin{equation}\label{eq:average_aoi}
    \bar{\Delta} = \sum_{a=1}^\infty a \,\pi(a) = \frac{1}{p\,Q}.
\end{equation}
Meanwhile, for this scenario, where the \gls{AoI} is always reset to $\Delta(t) = 1$ \changed{after the packet is received}, the average \gls{PAoI} is calculated by obtaining the mean recurrence time of state $\Delta(t) = 1$. As the \gls{DTMC} is irreducible, this is given by the inverse of the probability of an \gls{ED} having $\Delta(t)=1$ \cite{Gomez:LAA:2010}, i.e., 
\begin{equation}\label{eq:peak_aoi}
    \bar{\Delta}^{(p)} = \frac{1}{\pi(1)} = \frac{1}{p\,Q}.
\end{equation}
Note that, if $q(a)$ is not constant and independent of $\Delta(t)$, which depends on the adopted protocol, \eqref{eq:average_aoi} and \eqref{eq:peak_aoi} will not necessarily be equal, since~\eqref{eq:pdf_aoi_eds} is no longer valid, and the final expressions must be reevaluated, although the methods to obtain them remain the same. In our setup, both the relays and the \gls{AP} can keep track of the \gls{AoI} of the \glspl{ED}, the former by updating as soon as it receives a packet, and the latter using the information broadcasted by the \gls{AP}.

\section{AoI Lower Bound in the Multi-Relay Scenario}\label{sec:LowerBound}

In this section, we present a lower bound for the \gls{AoI} in a multi-relay scenario with ALOHA in the first phase, which serves as an optimal benchmark for the schemes evaluated in the next section. To this end, we expand on the formulation from~\cite{deJesus:EuCNC:2023} and consider the \gls{DTMC} model from~\cite{Chen:ICCC:2020}, providing expressions to calculate the \gls{AAoI}, \gls{PAoI}, and an approximation of the probability distribution of the \gls{AAoI}. The lower bound assumes that phase-2 is error and collision-free, {\it i.e.}, that all the packets received by the relays are correctly delivered to the \gls{AP} in the same time slot they are received. \changed{Thus, this analysis is equivalent to that of a single-hop ALOHA system with multiple \glspl{AP}, where packets can be captured in multiple \glspl{AP}, and collision events are independent at each \gls{AP}.}

Using~\eqref{eq:average_aoi} and~\eqref{eq:peak_aoi}, we can determine a lower bound for the \gls{AAoI} and \gls{PAoI} by properly defining an expression for $Q$. We focus on a single \gls{ED} and determine its probability of reaching one relay without being erased or colliding with other \glspl{ED}. The probability of $n$ other \glspl{ED} to be simultaneously active with the \gls{ED} of interest is
\begin{equation}
    P_n = \binom{N-1}{n}p^n(1-p)^{N-1-n}.
\end{equation} 

Let $u\leq n$ denote the number of \glspl{ED} in the same channel as the target \gls{ED}, which happens with probability
\begin{equation}
    P_u = \binom{n}{u}\frac{1}{F^u}\left(1-\frac{1}{F}\right)^{n-u}.
\end{equation}The \gls{ED} will have its packet collected at a relay with probability $\Tilde{Q}_u = (1-\varepsilon_1)\varepsilon_1^{u}$. For this packet to be received at the \gls{AP}, it needs to be captured by at least one relay, which happens with probability $1-(1-\Tilde{Q}_u)^K$. The final expression for $Q$ is given by adding this across all possible values of $n$ and $u$, multiplied by their respective probability of occurrence:
\begin{equation}\label{eq:Q}
    Q = \sum_{n=0}^{N-1}P_n\sum_{u=0}^nP_u(1-(1-\Tilde{Q}_u)^K),
\end{equation}
and the \gls{AAoI} and \gls{PAoI} are given by \eqref{eq:average_aoi} and \eqref{eq:peak_aoi}, respectively, resulting in
\begin{equation}\label{eq:to_minimize}
     \bar{\Delta} = \bar{\Delta}^{(p)} = \left({p\sum_{n=0}^{N-1}P_n\sum_{u=0}^nP_u(1-(1-\Tilde{Q}_u)^K)}\right)^{-1}.
\end{equation}

Along with the \gls{AAoI} and the \gls{PAoI}, we evaluate the performance of the system by showing the probability of the \gls{AAoI} to be bigger than a given value $\delta$, for any time slot. Let $\Delta$ be a random variable that captures the instantaneous average \gls{AoI} of the network, and let $\Delta_i$ be the random variables of the instantaneous \gls{AoI} of \glspl{ED}. Then, 
\begin{equation}
    {\Delta}(t) = \frac{1}{N}\sum_{i=1}^N \Delta_i(t),
\end{equation}
with the probability distribution of $\Delta$ given by \eqref{eq:pdf_aoi_eds}. 

\begin{theorem} \label{th:lb}
The instantaneous average \gls{AoI} of the network $\Delta(t)$ has probability distribution function $\mathbf{\Pi}(\delta)$ that is approximated by
\begin{equation}\label{eq:theorem}
    \mathbf{\Pi}(\delta) \approx  {P^N}\,\frac{(1-p\,Q)^{\delta N}}{(N-1)!}\frac{\Gamma(\delta N)}{\Gamma(\delta N - N + 1)},
\end{equation}
where $P\triangleq{p\,Q}/({1-p\,Q})$.
\end{theorem}
\begin{IEEEproof}
    See Appendix \ref{app:proof}.
\end{IEEEproof}

With this result in hand, one can evaluate the system's performance by taking the complement of the cumulative distribution of $\Delta$, obtaining the fraction of time slots where the \gls{AAoI} is above a certain value $\delta$.  

\section{Max-Age Algorithms}\label{sec:MAA}

In \cite{Tripathi:GLOBECOM:2017}, the authors propose the \gls{MAM} and \gls{IMAS} algorithms to a single-hop multiple access network. In a network operating under the aforementioned algorithms, the \gls{AP} is responsible for properly assigning the channels to the \glspl{ED} with the highest \gls{AoI}, aiming at minimizing the network \gls{AAoI}. Although both algorithms are asymptotically optimal, \gls{MAM} outperforms \gls{IMAS} at the cost of much higher algorithmic complexity and equivalent information exchange requirements.

In the \gls{MAM} algorithm, the set of \glspl{ED} and channels are considered as opposite vertices of a bipartite graph, such that a given \gls{ED}-channel pair is connected by an edge in case the corresponding link is not erased, and the instantaneous \gls{AoI} of the \glspl{ED} is used as the weight of each edge. The graph is used as the input of a matching algorithm (\textit{e.g.}, the Hungarian Algorithm~\cite{Kuhn:NRLQ:1955}), which is responsible for finding the max-weight matching in this weighted bipartite graph, providing the optimal channel allocation in that time slot that maximizes the sum of the \gls{AoI} of the users transmitting.

In~\gls{IMAS}, starting from the first channel, the algorithm searches for the \gls{ED} with the highest \gls{AoI} whose link to the considered channel is not erased in the time slot, allocating such a channel to the \gls{ED}. The \gls{ED} is removed from the contest and the process is repeated until there are no channels or connectable \glspl{ED} left.

\subsection{Max-Age Algorithms in Multi-Relay Scenario}
\begin{algorithm}[tb]
\caption{Max-Age Matching Multi-Relay}\label{alg1}
 \hspace*{\algorithmicindent} \textbf{Input:} Connectivity matrix $\mathbf{H}$; Index of users from  
 \hphantom{\hspace*{\algorithmicindent} \textbf{Input: }} packets decoded by relays $\mathcal{R}$; Current AoI\\
 \hphantom{\hspace*{\algorithmicindent} \textbf{Input: }}of \glspl{ED} $\{\Delta_i(t)\}_{i=1}^N$;\\
 \hspace*{\algorithmicindent} \textbf{Output:}  Allocation of channels to relays $\mathcal{F}^\star$ 
\begin{algorithmic}[1]
\State $\mathcal{S} \gets \{S_j\}_{j=1}^{|\mathcal{S}|} = \texttt{UniqueCombinations}(\mathcal{R})$ \label{alg1:UniqueCombinations}
\State $\text{AoISum} \gets 0$
\State $j \gets 1$
\While{$j \leq |\mathcal{S}|$}
    \State $E_j \gets \texttt{AdjacencyMatrix}(S_j, \mathcal{F}, \mathbf{H})$ \label{alg1:AdjacencyMatrix}
    \State $\mathcal{G}_j \gets \texttt{Graph}(S_j, \mathcal{F},E_j)$ \label{alg1:Graph}
    \State $\hat{\mathcal{F}}_j \gets \texttt{MaxAgeMatching}(\mathcal{G}_j,\{\Delta_i(t)\}_{i=1}^N)$ \label{alg1:MaxAgeMatching}
    \State $\text{AoISum}_j \gets \sum_{S_j} \mathcal{A}$
    \If{$\text{AoISum}_j > \text{AoISum}$}
        \State $\mathcal{F}^\star \gets \hat{\mathcal{F}}_j$ 
        \State $\text{AoISum} \gets \text{AoISum}_j$ 
    \EndIf
    \State $j \gets j + 1$
\EndWhile
\end{algorithmic}
\end{algorithm}

In this work, we extend the single-hop max-age algorithms to the two-hop case. In the single-hop scenario, the set of \glspl{ED} competing for a channel is connected to edges with weights that are independent of the weights of the other edges. In the two-hop case considered in this work, since two or more relays may have the same packet to transmit, these algorithms must be modified to address this replication, so that resources are not used for redundant information. In fact, if the \gls{ED} with the highest instantaneous AoI is received by more than one relay, all such relays would have the highest priority to forward the packet when following these max-age algorithms, compromising the overall performance. To solve this issue, we present the modified versions of \gls{MAM} and \gls{IMAS} in Algorithms~\ref{alg1} and~\ref{alg2}, respectively. 

In Algorithm~\ref{alg1}, we describe the Max-Age Matching algorithm adapted to the multi-relay scenario. It requires the connectivity matrix, which relates the relays to the channels, the index of the users whose packets were captured by the relays, and the \gls{AoI} of these users. In line \ref{alg1:UniqueCombinations}, the function $\texttt{UniqueCombinations}(\cdot)$ takes the argument $\mathcal{R}$ and generates sets $S_j$ of up to $F$ elements, the number of channels available. Each element in the set is the index of a relay, such that no two relays have captured the same packet. These sets are considered as the vertices of a graph, together with the vertices of the channels, denoted by $\mathcal{F}$. Then, for each set generated, the algorithm performs the following. First, it generates an adjacency matrix in line~\ref{alg1:AdjacencyMatrix}, using the information from the connectivity matrix $\mathbf{H}$ to determine the edges of a graph. Then, in line~\ref{alg1:Graph}, a bipartite graph is generated, with $S_j$ and $\mathcal{F}$ as the partitions of the graph and $E_j$ as the edges connecting these partitions. This graph and the \gls{AoI} vector are used in line~\ref{alg1:MaxAgeMatching} as inputs of the Max-Age Matching algorithm, as described earlier. Then, the sum of the resulting \gls{AoI} is compared to the highest value of the previous $j-1$ iterations (with the sum set to $0$ for $j=0$), and the current channel allocation is considered as the final allocation if the sum is the highest so far. \changed{Note that, as was the case for the single-hop setting, in the two-hop implementation,  only the channels that are not erased are considered for generating the adjacency matrix. Thus, this algorithm avoids collisions and only allows the transmission of packets that will not be erased.}
\begin{algorithm}[tb]
\caption{Iterative Max-Age Scheduling Multi-Relay}\label{alg2}
  \hspace*{\algorithmicindent} \textbf{Input:} Connectivity matrix $\mathbf{H}$; Index of users from
 \hphantom{\hspace*{\algorithmicindent} \textbf{Input: }} packets decoded by relays $\mathcal{R}$; \gls{AoI} of \glspl{ED} \\
\hphantom{\hspace*{\algorithmicindent} \textbf{Input: }}  $\{\Delta_i(t)\}_{i=1}^N$; \\
\hspace*{\algorithmicindent} \textbf{Output:} Allocation of channels and power levels to\\
\hphantom{\hspace*{\algorithmicindent} \textbf{Output: }}relays $\mathcal{F}^\star$ 
\begin{algorithmic}[1]
\State $\mathbf{h}_f \gets$ columns of matrix $\mathbf{H}$, for $f=\{1,2,\dots F\}$
\State $f \gets 1$
\While{$f \leq F$}
    \State $({u}, {k}) \gets \texttt{Max}(\mathcal{R}, \{\Delta_i(t)\}_{i=1}^N, \mathbf{h}_f)$  \label{alg2:Max}
    \State $\mathcal{R} \gets \mathcal{R}\backslash \{u\}$
    \State $\mathcal{F}^\star(f) \gets k$
    \State $f \gets f + 1$
\EndWhile
\end{algorithmic}
\end{algorithm}

In Algorithm \ref{alg2}, we describe the Iterative Max-Age Scheduling algorithm adapted to the multi-relay scenario. As in the multi-relay version of the \gls{MAM} algorithm, it requires the connectivity matrix, which relates the relays to the channels, the index of the users whose packets were captured by the relays, and the \gls{AoI} of these users. However, this algorithm is much less complex, with complexity $O(KF)$, compared to $O(K^3)$ of \gls{MAM} \cite{Tripathi:GLOBECOM:2017}. For each channel $f$, the algorithm performs the following. In line \ref{alg2:Max}, the arguments $\mathbf{H}$, $\mathcal{R}$ and $\{\Delta_i(t)\}_{i=1}^N$ are taken by the function $\texttt{Max}(\cdot)$, which returns the \gls{ED} with the highest \gls{AoI}, given that at least one relay that has captured its packets has a viable connection to channel $f$. Then, any replicas of the packet from the selected \gls{ED} are removed from the list of packets to be forwarded, and the $f$-th channel is allocated to the selected relay. 

However, both MAM and IMAS algorithms run in a centralized manner at \gls{AP}.  This requires a large amount of information exchange between relays and \gls{AP}, which may be prohibitive in practice. Thus, in what follows, we propose two algorithms in which the allocation of channels to the relays is defined in a decentralized manner, only demanding the supervision of the \gls{AP}.

\section{Proposed Age-Based Delayed Request Schemes}\label{sec:ABDR}

In this section, we propose an alternative to the aforementioned max-age algorithms that aims to reduce information exchange without compromising network performance, which we refer to as \gls{ABDR}. It is worth mentioning that, similarly to \gls{MAM} and \gls{IMAS}, \changed{we consider that, if relays successfully receive the pilot sequence in a channel, meaning it is not erased, all the signaling exchange in that channel between relays and \gls{AP} happens error-free.} 

\subsection{Age-Based Delayed Request}
\changed{
In \gls{ABDR}, the phase-2 transmission slot is divided into four sub-slots. }
\begin{itemize}
    \changed{\item In sub-slot 1, the \gls{AP} broadcasts a pilot signal in all channels, using the same transmit power relays would use. Assuming channel reciprocity, the relays determine whether or not their channels to the \gls{AP} are erased based on the successful reception of the pilot signal.
    \item In sub-slot 2, the relays that have packets to transmit and whose channel is not erased schedule their \gls{RTS}, similarly to~\cite{Bletsas:ITWC:2007}. To this end, the relay starts a timer that expires in a time that is inversely proportional to the ratio between the \gls{AoI} of the user whose packet they collected, and the highest \gls{AoI} among devices. To avoid multiple relays starting their \gls{RTS} simultaneously, a random variable is introduced. Let $\Delta_{f,k}$ denote the \gls{AoI} of the user captured by the $k$-th relay in channel $f$, and let $\Delta_{\text{max}}$ represent the highest \gls{AoI} among all the $N$ users. Then, the timer set by the $k$-th relay in the $f$-th channel will expire at time \begin{equation}\label{eq:tau_fk}
    \tau_{f,k} = \min\{1 - \Delta_{f,k} / \Delta_{\text{max}} + \tau, 1\},
    \end{equation}
    where {$\tau \sim U(0,t^\star)$} is a random variable included so that there are no collisions between relays that captured packets from \glspl{ED} with the same \gls{AoI} (or replicated packets from the same \gls{ED}). The value of $t^\star$ can be selected depending on the processing times of relays. Note that, since packets are restricted to being forwarded in the same channels they were captured, no redundant information is transmitted. Furthermore, a relay can only transmit in a single channel at each slot, canceling the requests in other channels if it was able to perform its \gls{RTS}. Once the timer is expired, if the channel was idle until then, the relay emits a signal and holds it until the end of sub-slot 2.  The \gls{RTS} procedure is presented in Algorithm~\ref{alg3}.
\item In sub-slot 3, the \gls{AP} sends a \gls{CTS} to the relay, reserving the channel for it. 
\item  Lastly, in sub-slot 4, the relay that received the \gls{CTS} transmits the packet it captured. Here, the packet is transmitted without the need for a pilot sequence, but only user identification is still required.}
\end{itemize}

\begin{algorithm}[tb]
\caption{ABDR - \gls{RTS} procedure, channel $f$ and relay $k$}\label{alg3}
 \hspace*{\algorithmicindent} \textbf{Input:} Status (erased or not) of channel $f$ for 
 \hphantom{\hspace*{\algorithmicindent} \textbf{Input: }}relay $k$, $h_{f,k}$; Index of user decoded by relay $k$ \\\hphantom{\hspace*{\algorithmicindent} \textbf{Input: }}in channel $f$, $r$; maximum random delay $t^\star$; \\\hphantom{\hspace*{\algorithmicindent} \textbf{Input: }}\gls{AoI} of \glspl{ED} $\{\Delta_i(t)\}_{i=1}^N$;\\
 \hspace*{\algorithmicindent} \textbf{Output:}  Timing of \gls{RTS} for relay $k$ in channel $f$, $\tau_{f,k}$ 
\begin{algorithmic}[1]
\State $\tau \gets \texttt{random}(0,t^\star)$
\State $\Delta_{f,k} \gets \Delta_r(t)$
\State $\Delta_{\text{max}} \gets \max(\{\Delta_i(t)\}_{i=1}^N)$
\If{$h_{f,k}$}
    \State $\tau_{f,k} \gets$ \eqref{eq:tau_fk}
\EndIf
\end{algorithmic}
\end{algorithm}

\subsection{Buffered ABDR}
The \gls{ABDR} scheme is limited in a few aspects. For instance, when multiple packets are collected only in $f<F$ channels, only at most $f$ of them will get the chance to be forwarded to the \gls{AP}, with the remaining packets being discarded. If, instead, these packets are buffered to the next time slots, they have a better chance of being forwarded, potentially reducing the average \gls{AoI} of the network.

In the \gls{B-ABDR} scheme, each relay has a buffer with capacity for $B$ packets, which is updated at the end of each time slot. At each time slot, the relay will calculate the request delay for all the packets it has captured in phase-1, alongside the packets it has stored in the buffer. The remainder of the transmission decision is as in \gls{ABDR}, including the restriction on the channels used to forward the packets.

A key difference from the buffered \gls{B-ABDR} to its memoryless counterpart is that packets can reach the \gls{AP} already aged. To account for this, the \gls{AP} considers the time-stamps in the packets it receives. Then, when it feed backs the index of the received packets, it also informs the actual age of the \glspl{ED} to which these packets belong. Lastly, relays use the information from \gls{AP} to discard packets already collected, as well as to only keep the $B$ packets with the highest \gls{AoI} in their buffers. 

One could also extend \gls{IMAS} and \gls{MAM} to a buffered version, although it incurs a further increase in signaling. The relays now need to have enough time allocated to transmit the packet id of up to $B$ packets, and the \gls{AP} needs to inform the relays which of the packets should be transmitted. The impact of this is discussed in the sequence for the buffered version of \gls{IMAS}, which we refer to as \gls{B-IMAS}. 

\subsection{Information Exchange Demanded by the Algorithms}

\begin{table*}[tb]
    \centering
    \caption{Time required for information exchange with the \gls{AP}.}
    \begin{tabular}{l c c c c c c c }
    \toprule
                           & Pilot Sequence & Packet ID   & Grant &  RTS  &   CTS     &      Packet Transmission   & Acknowledge   \\
    \midrule
         IMAS/MAM          & $K T_p$        & $K T_i$     & $T_k$ &     -  &    -       &   $T-K T_p - K T_i - 2T_k$   & $T_i$  \\
         Buffered IMAS/MAM & $K T_p$        & $K B T_i$   & $T_k + T_i$ &     -  &  -         &   $T-K T_p - K B T_i - 2T_k$    & $T_i$  \\
         B-/ABDR           & $T_p$          &    -         &  -    & $T_r$ & $T_k$ &   $T - T_p - T_r - 2T_k$ & $T_i$ \\
         \bottomrule
    \end{tabular}
    \label{tab:my_label}
\end{table*}

We compare the volume of information exchange by evaluating the time required to transmit each piece of information for each algorithm. Let $T$ be the length of the phase-2 transmission slot in symbols. The pilot sequence transmitted by the relays (\gls{IMAS} and \gls{MAM}) or the \gls{AP} (\gls{ABDR} and \gls{B-ABDR}) is denoted by $T_p$ and assumed to be of the same length in all cases. The length of the packet ID is denoted by $T_i$ and should not take more bits than $\lceil \log_2 N\rceil$. Moreover, the minimal information required for the grant and for the \gls{CTS} is the relay ID, denoted by $T_k$. Lastly, the \gls{RTS} signal is $T_r$ symbols long, and the acknowledge has the same length of the grant signal, $T_k$ symbols. A table comparing the amount of time for information exchange per time slot is summarized in Table~\ref{tab:my_label}. 

From Table \ref{tab:my_label}, one can derive the maximum duration of $T_r$ so that \gls{ABDR} and \gls{B-ABDR} are still advantageous. When compared to memoryless \gls{IMAS} and \gls{MAM}, $T_r < (K-1)T_p + KT_i$. On the other hand, for the buffered versions of \gls{IMAS} and \gls{MAM}, $T_r < (K-1)T_p + K(B+1)T_i +
T_i$. It is desirable that the length of $T_r$ is long enough to avoid collisions in imperfect timers but should still be shorter than the sum of the other signaling. 

\begin{figure}[!t]
    \centering
    \includegraphics[width=\linewidth]{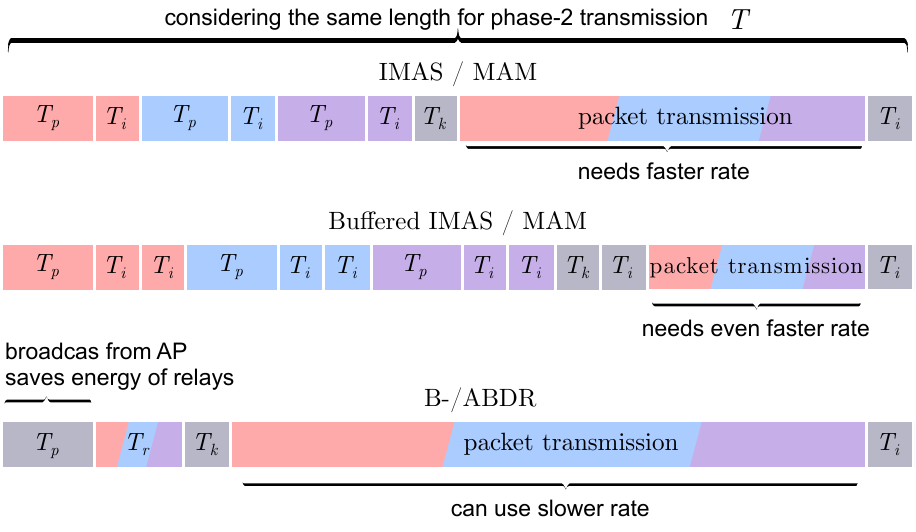}
    \caption{Timing of the information exchange in phase-2 of transmission sub-slot for a channel. For illustration purposes, we consider $K=3$ relays and a buffer size of $B=1$ packets. }
    \label{fig:TimingScheme}
\end{figure}

In Fig.~\ref{fig:TimingScheme}, we illustrate this for the case with $K=3$ and $B=1$. The colored blocks refer to relays transmitting to the \gls{AP}, and the gray blocks refer to the \gls{AP} transmitting to relays. This representation shows another advantage of the proposed method. Not only is the required volume of signaling much smaller than \gls{IMAS} and \gls{MAM}, but the volume also does not scale with $K$ and $B$, allowing the inclusion of more relays and memory space if necessary. \changed{Furthermore, as \gls{ABDR} and \gls{B-ABDR} require less signaling, a larger portion of the phase-2 transmission slot can be reserved for the packet transmission, allowing for larger packets to be transmitted, or for a lower rate to be used, increasing reliability.} \changed{Nevertheless, as relays do not need to transmit metadata, the energy consumption by these devices in the signaling phase can be significantly decreased.}

For the case with $K=5$ and $B=1$, let us consider a numerical example with $N=30$. For simplicity, let us consider a BPSK modulation with each symbol corresponding to one bit. We assume the pilot length to be $T_p=5$ symbols, $T_i=\lceil \log_2 N\rceil = 5$ symbols and $T_k=\lceil \log_2 K\rceil = 3$ symbols. Moreover, consider a typical {IoT} packet of 20~B, resulting in 160 symbols. Thus, under IMAS and MAM the entire phase-2 must fit $T=173$ symbols. For the buffered versions, this grows to $T=203$ symbols, requiring a faster rate so that it fits in the same total time as the other approaches for a fair comparison, which is more prone to errors. On the other hand, if ABDR and B-ABDR use the same rate as IMAS and MAM, the RTS phase can last for $T_r=45$ symbols. For the rate used by the buffered versions of IMAS and MAM, the RTS phase can last for $T_r=75$ symbols. Suppose this number of symbols is more than enough for performing the RTS, depending on the resolution of the timers of relays. In that case, the RTS phase of ABDR/B-ABDR can be shorter than the signaling of IMAS/MAM, and the rate used for the transmission in ABDR/B-ABDR can be decreased, favoring implementation and reliability. 

\changed{For the case with $K=5$ and $B=1$, let us consider a numerical example with $N=30$. For simplicity, let us consider a BPSK modulation with each symbol corresponding to one bit. We assume the pilot length to be $T_p=5$ symbols, $T_i=\lceil \log_2 N\rceil = 5$ symbols and $T_k=\lceil \log_2 K\rceil = 3$ symbols. Moreover, consider a typical \gls{IoT} packet of 20~Bytes, resulting in 160 symbols. Thus, under IMAS and MAM the entire phase-2 must fit $T=5(5+5)+3+160+5=218$ symbols. For the buffered versions, this grows to $T=248$ bytes, requiring a faster rate, which is more prone to errors. On the other hand, if ABDR and B-ABDR use the same rate as IMAS and MAM, the RTS phase can last for $T_r=45$ symbols. For the rate used by the buffered versions of IMAS and MAM, the RTS phase can last for $T_r=75$ symbols. Suppose this number of symbols is more than enough for performing the RTS, depending on the resolution of the timers of relays. In that case, the RTS phase can be shorter, and the rate used for the transmission in ABDR/B-ABDR can be decreased, favoring implementation and reliability. }

\changed{The multi-relay versions of the IMAS and MAM algorithms both require a significant overhead load due to how they were originally proposed. Our proposed schemes significantly decrease this load while still prioritizing packets from high-\gls{AoI} users. The resulting overhead load is close to the minimal required in typical setups (pilot transmission and acknowledgment) apart from pure ALOHA implementations.}

\changed{
\subsection{Computational Complexity and Memory Overhead of the Algorithms}
The \gls{MAM}'s and \gls{IMAS}' computational complexity was studied in the single-hop case in \cite{Tripathi:GLOBECOM:2017}, depending only on the number of \glspl{ED} (\gls{MAM}) and also on the number of channels (\gls{IMAS}). For the two-hop scenario, the complexity of \gls{MAM} and \gls{IMAS} is slightly different, with complexity of $\mathcal{O}(K^3)$ and $\mathcal{O}(FK)$, respectively.

On the other hand, the complexity of \gls{ABDR} and \gls{B-ABDR} is moved to the relays, instead of the \gls{AP}. While this would typically be undesirable, it is justifiable by the significantly lower complexity of the two proposed algorithms. For \gls{ABDR}, for instance no looping is necessary, and both division and random number generation operations incur a complexity of $\mathcal{O}(1)$. On the other hand, \gls{B-ABDR} requires a sorting at each time slot, and the computational complexity is $\mathcal{O}(B+1)$ if considering, for example, the bubble sort implementation. 

Regardless of its effectiveness in guaranteeing low-\gls{AoI} for the network, \gls{B-IMAS} comes at the cost of increased memory requirements compared to the other algorithms. Although the relays are assumed to be devices with more computational and memory resources compared to the \glspl{ED}, these resources are not unlimited. Thus, the size of the buffer $B$ is limited, and the relay must employ a memory control to avoid overflowing the buffer. For instance, we consider that the packets belonging to users with high \gls{AoI} are prioritized over packets from users with low \gls{AoI}, and the ($B+1$)-th oldest packet is discarded at the end of each time slot, but other approaches can be implemented depending on the application.  
}

\section{Numerical Results}\label{sec:results}
In this section, we present some simulation results to evaluate the performance of the proposed algorithms. As a baseline, we also present the results for the lower-bound developed in Section~\ref{sec:LowerBound}, with simulation results to validate our approximations. The theoretical results are exact regarding the \gls{AAoI} and \gls{PAoI} of the lower-bound, but are approximations for the \gls{AAoI} \gls{CDF}. In the last column of Table~\ref{tab:list_of_symbols}, we summarize the default simulation parameters used throughout our simulations, unless stated otherwise. In Figs.~\ref{fig:AAoI_relays_1st}, \ref{fig:AoI_relays}, \ref{fig:AAoI_Fp}, and~\ref{fig:erasure}, continuous lines with specific markers (square, circle, diamond, and so on) represent \gls{AAoI} obtained with simulations, while the tips of the bars that emerge from the markers represent the \gls{PAoI}. In all figures, the red asterisks represent the results for the lower bound obtained with the analytical formulation.

\subsection{Performance Improvement with Information Exchange}
\begin{figure}
    \centering
    \includegraphics[width=0.5\textwidth]{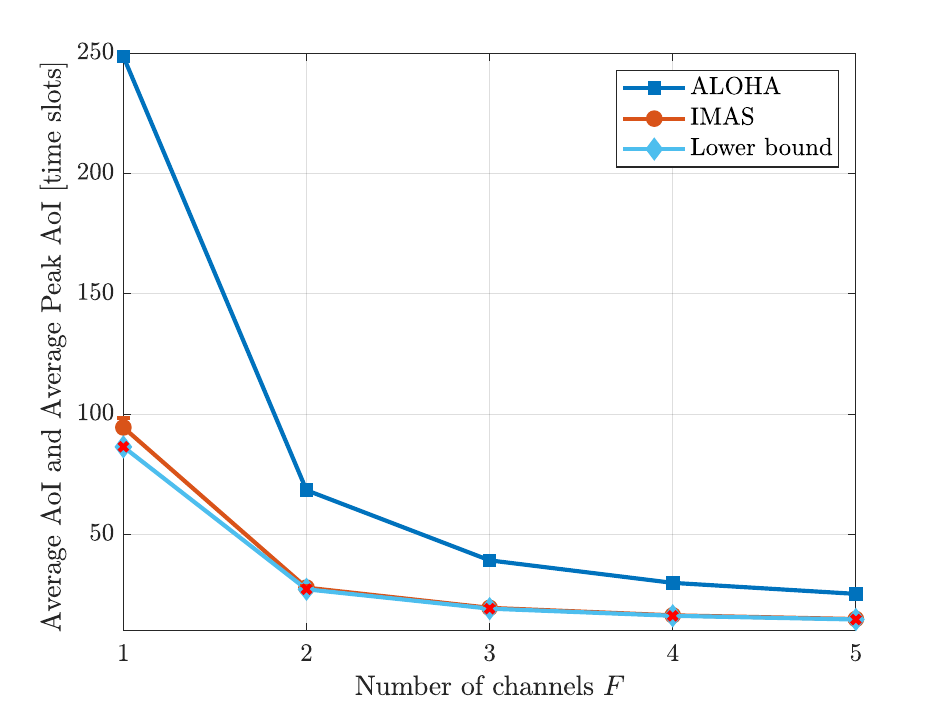}
    \caption{\changed{The \gls{AAoI} and \gls{PAoI} as a function of $F$ for the {ALOHA} and \gls{IMAS} protocols, and for the lower-bound. This highlights the effectiveness of the coordinated control of relays in achieving the lower bound.}}
    \label{fig:AAoI_relays_1st}
\end{figure}

{First, we illustrate the benefits of performing an exchange phase before relays forward the packets they have received. To this end, we present the performance gain in terms of \gls{AAoI} and \gls{PAoI} obtained with \gls{IMAS} compared to the classical ALOHA scheme, as the number of frequency channels increases from $F=1$ to $F=5$, in Fig. \ref{fig:AAoI_relays_1st}. These results show that, although this exchange comes with a cost, the performance gain justifies it, as, with as few as $F=1$ channels, the performance is already significantly close to optimal, and with $F\geq2$ the performance of \gls{IMAS} is almost indistinguishable from that of the lower bound.}

\subsection{The Number of Relays}
\begin{figure}[tp]
    \centering
    \includegraphics[width=0.5\textwidth]{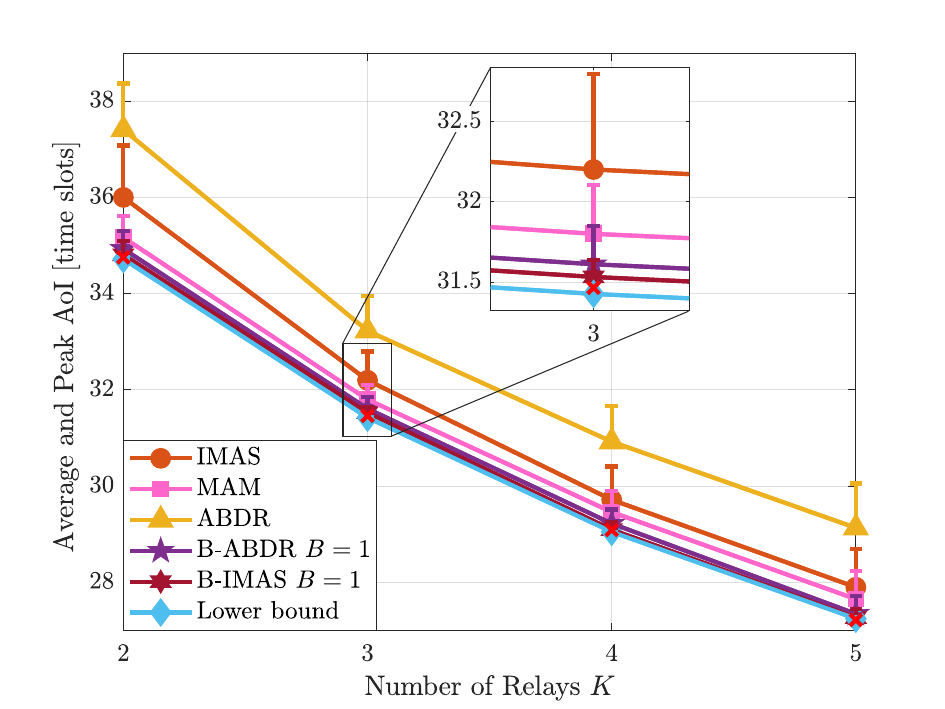}
    \caption{\changed{The \gls{AAoI} and \gls{PAoI} as a function of the number of relays $K$. Overall, all algorithms scale well with the number of relays, though \gls{MAM} and \gls{IMAS}, and more severely \gls{B-IMAS}, require proportionally more time dedicated to signaling.}}
    \label{fig:AoI_relays}
\end{figure}

Herein, we focus on the algorithms that have information exchange with the \gls{AP} in the second hop, namely, \gls{IMAS}, \gls{MAM}, \gls{B-IMAS}, \gls{ABDR}, and \gls{B-ABDR}, compared to the lower-bound, by assessing the impact of the number of relays in their performance. In Fig. \ref{fig:AoI_relays}, we present the \gls{AAoI} and the \gls{PAoI} achieved by each method when the number of relays varies from $K=2$ to $K=5$. 
The first thing to notice is that both \gls{IMAS} and \gls{MAM} have a performance gap for \gls{AAoI} and \gls{PAoI} that is less than one time slot in this setup, contrary to the single-hop case where this gap is more pronounced. In the single-hop case, the increase in algorithmic complexity of the latter is justified by the performance gains, which is not the case in a two-hop case. Moreover, our proposed \gls{ABDR} algorithm has a performance comparable to \gls{IMAS} and \gls{MAM}, although with a significantly lower volume of information exchange and algorithmic complexity, even as the number of relays increases. Meanwhile, \gls{B-ABDR} is even more competitive by outperforming all other algorithms, except for \gls{B-IMAS},  with a buffer size as small as $B=1$ packet, while only requiring the same volume of information exchange of that of \gls{ABDR}. On the other hand, \gls{B-IMAS} performs slightly better than \gls{B-ABDR}, with a significant cost in information exchange, as discussed earlier, while both are extremely close to the lower-bound, which gives \gls{B-ABDR} a clear advantage.  The evaluation of the \gls{PAoI} further confirms that the \gls{ABDR} scheme is competitive with the more complex schemes, and that the \gls{B-ABDR} scheme is the superior algorithm in resource utilization and competitive \gls{AoI} performance.

For brevity,  we do not include results of \gls{MAM} and \gls{B-IMAS} hereinafter. This is justified by the fact that \gls{MAM} and \gls{IMAS} have similar performances, and that \gls{B-IMAS} has a similar performance compared to \gls{B-ABDR}, but with prohibitively large volumes of information exchange.

\subsection{Optimal Activation Probability}

\begin{figure}[tp]
    \centering
    \includegraphics[width=0.5\textwidth]{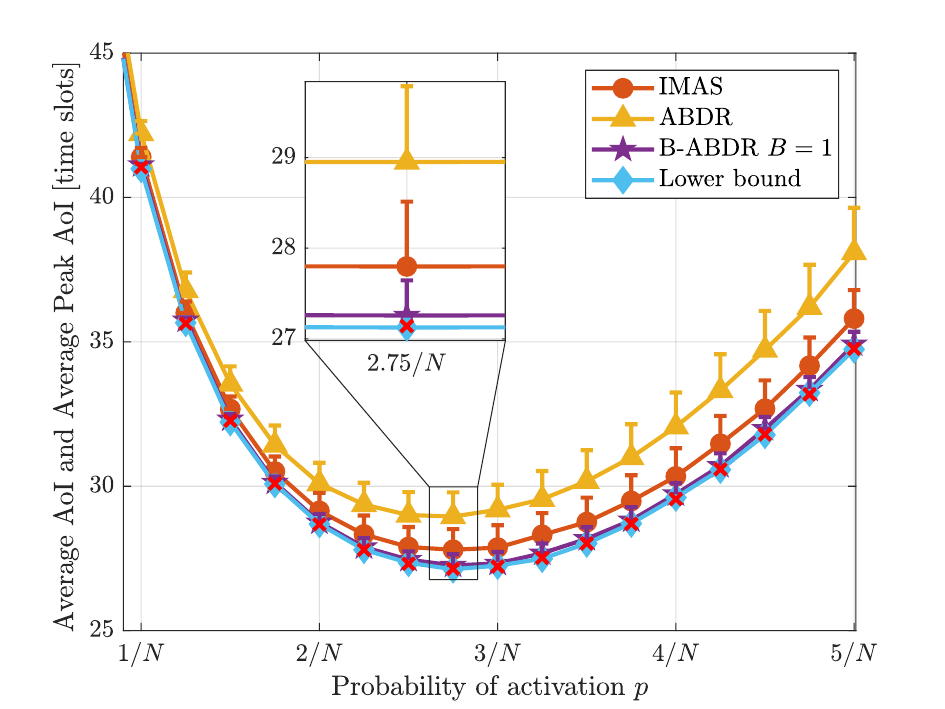}
    \caption{\changed{\gls{AAoI} and \gls{PAoI} as a function of $p$. All algorithms achieve similar performance and a minimal at the same value of $p$, allowing for the numerical optimization of $p$ by using the expressions provided.}}
    \label{fig:AAoI_p}
\end{figure}

\begin{figure}[tp]
    \centering
    \includegraphics[width=0.5\textwidth]{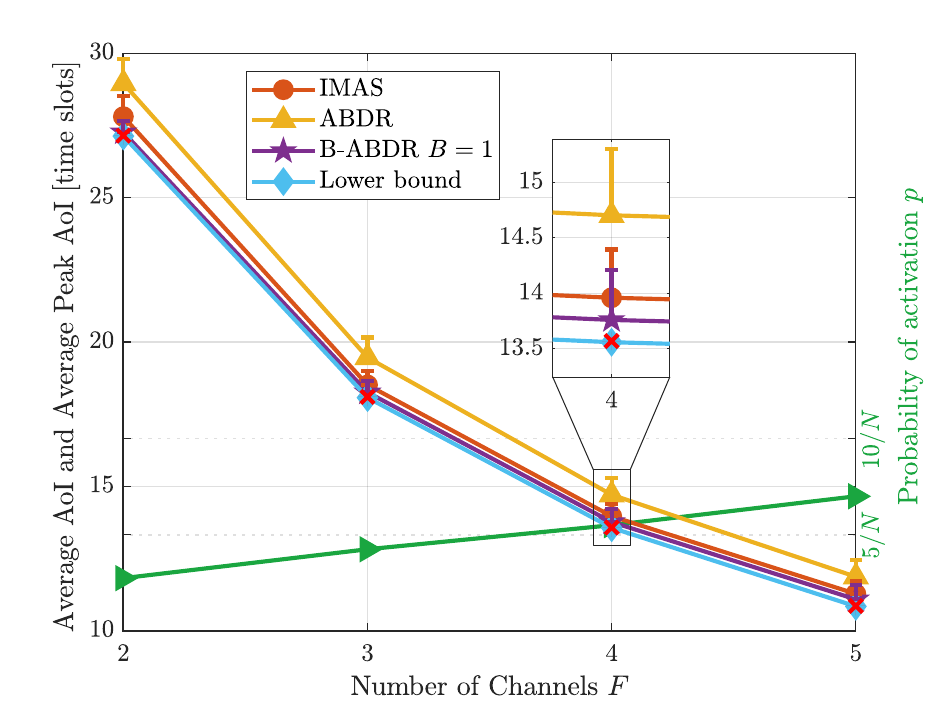}
    \caption{\changed{The \gls{AAoI} and \gls{PAoI} as a function of $F$ and for optimized values of $p$, and the optimal values of $p$. The numerical optimization is shown to be effective even as the network parameters change.}}
    \label{fig:AAoI_Fp}
\end{figure}

As phase-1 is ALOHA-based, there exists an optimal value for the activation probability that optimizes the \gls{AoI} performance. In Fig.~\ref{fig:AAoI_p}, we present the \gls{AAoI} performance of the discussed algorithms as a function of the activation probability $p$ for a scenario with $K=5$ relays and $F=2$ channels. Indeed, in this case, all algorithms have a similar trend, achieving a minimal \gls{AAoI} at $p=0.0917$. Thus, one can numerically minimize \eqref{eq:average_aoi} to obtain the activation probability that guarantees the best performance achievable by the phase-2 algorithms. To highlight this, in Fig. \ref{fig:AAoI_Fp}, we present the \gls{AAoI} and \gls{PAoI} of the discussed algorithms as a function of the number of channels $F$ when $p$ is optimized relying on \eqref{eq:average_aoi}. Overlayed with the \gls{AAoI} and \gls{PAoI}, we present the optimal values of $p$ that we have obtained.

\subsection{Average of the Instantaneous \gls{AoI} of \glspl{ED}}
\begin{figure}
    \centering
    \includegraphics[width=0.5\textwidth]{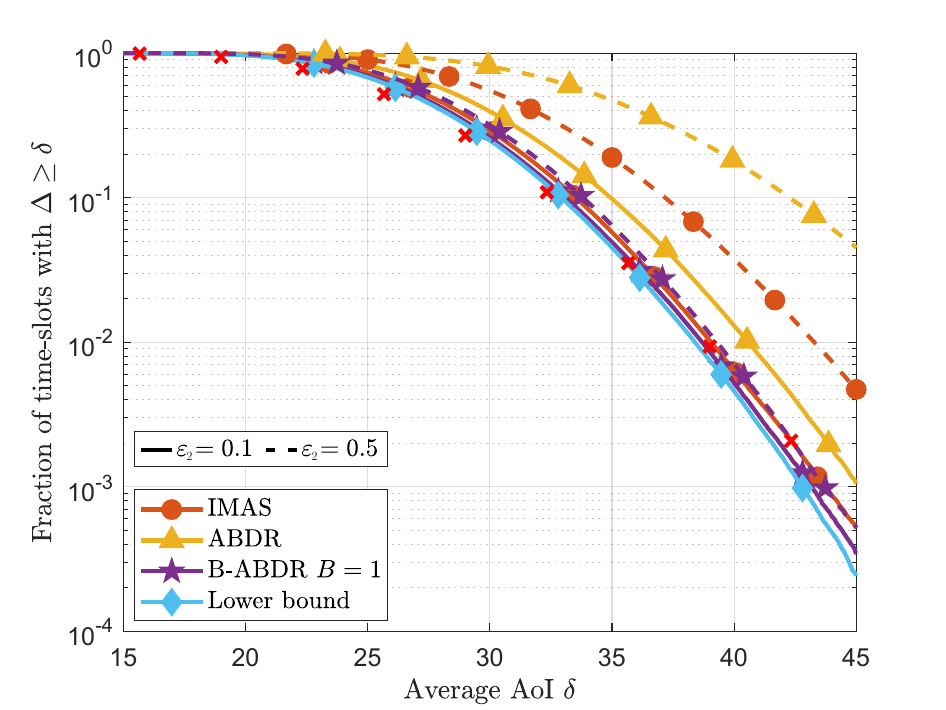}
    \caption{\changed{\gls{CDF} of the complement of the average instantaneous \gls{AoI} of \glspl{ED}. \gls{B-ABDR} achieves results close to the bound even at high erasure rates in phase-2. }}
    \label{fig:cost}
\end{figure}

In Fig. \ref{fig:cost}, we present the fraction of time slots where the \gls{AAoI} is above a value $\delta$. Here, we present results for erasure rate in the second phase of $\varepsilon_2=0.1$, as before, and also a scenario with a worse connectivity performance, setting $\varepsilon_2=0.5$, while $\varepsilon_1=0.1$. For $\varepsilon_2=0.1$, \gls{IMAS} and \gls{B-ABDR} have a similar performance, close to optimal, while \gls{ABDR} falls shorter. For the case with worse connectivity, the performance of \gls{IMAS} deteriorates significantly, while \gls{B-ABDR} still guarantees great performance, comparable to the case with $\varepsilon_2=0.1$.

\subsection{Impact of phase-2 erasure rate}
\begin{figure}
    \centering
    \includegraphics[width=0.5\textwidth]{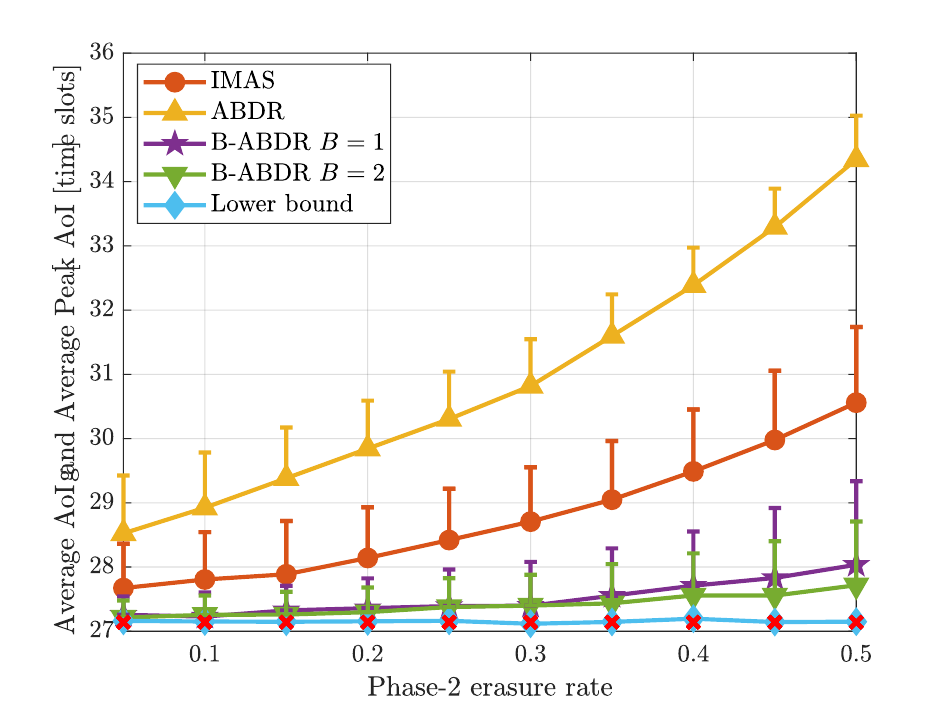}
    \caption{\changed{\gls{AAoI} as a function of the erasure rate in phase-2 $\varepsilon_2$. While the performance of the other algorithms quickly diverge from the lower-bound, \gls{B-IMAS} has a performance gap of at most one time slot in \gls{AAoI}, and two in \gls{PAoI}, even in its simplest configuration of $B=1$.}}
    \label{fig:erasure}
\end{figure}

In Fig.~\ref{fig:erasure}, we evaluate the impact of the increase in the erasure rate in phase-2 for \gls{IMAS}, \gls{ABDR}, and \gls{B-ABDR}. Compared to the other schemes, \gls{B-ABDR} is far more robust to the increase in the erasure rate, with performance close to optimal even in the worse erasure rate considered. This behavior highlights the main advantages of \gls{B-ABDR} compared to \gls{IMAS}. The buffered packets offer degrees of freedom, allowing the relays to explore favorable realizations of the channel to transmit packets. This guarantees that the \gls{AP} has fresh information from all \glspl{ED}, even if delayed by a few time slots, all while requiring minimal information exchange prior to packet transmission.

\changed{\subsection{Scalability of the proposed method}
In Fig.~\ref{fig:number_of_EDs}, we show how our proposed scheme scales with the increase in the number of \glspl{ED}. We vary $N$ from $30$ to $300$ and present the \gls{AAoI} and \gls{PAoI} as a function of $N$. As $N$ increases, the performance gap between \gls{ABDR}, \gls{IMAS} and \gls{B-ABDR} also increases, but \gls{B-ABDR} performance is kept significantly close to the lower bound even at $B=1$ packets buffered in each relay. 
\begin{figure}
    \centering
    \includegraphics[width=\linewidth]{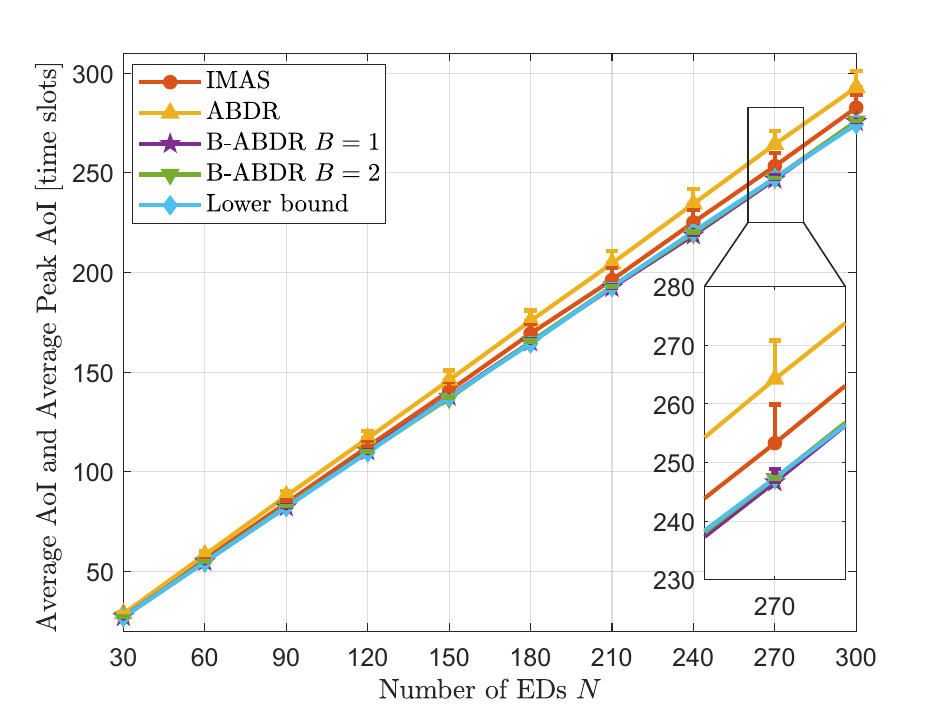}
    \caption{\changed{AAoI and PAoI as a function of the number of EDs. B-ABDR is the algorithm that better approaches the lower bound even at high counts of \glspl{ED}.}}
    \label{fig:number_of_EDs}
\end{figure}
}

\begin{figure}
    \centering
    \includegraphics[width=\linewidth]{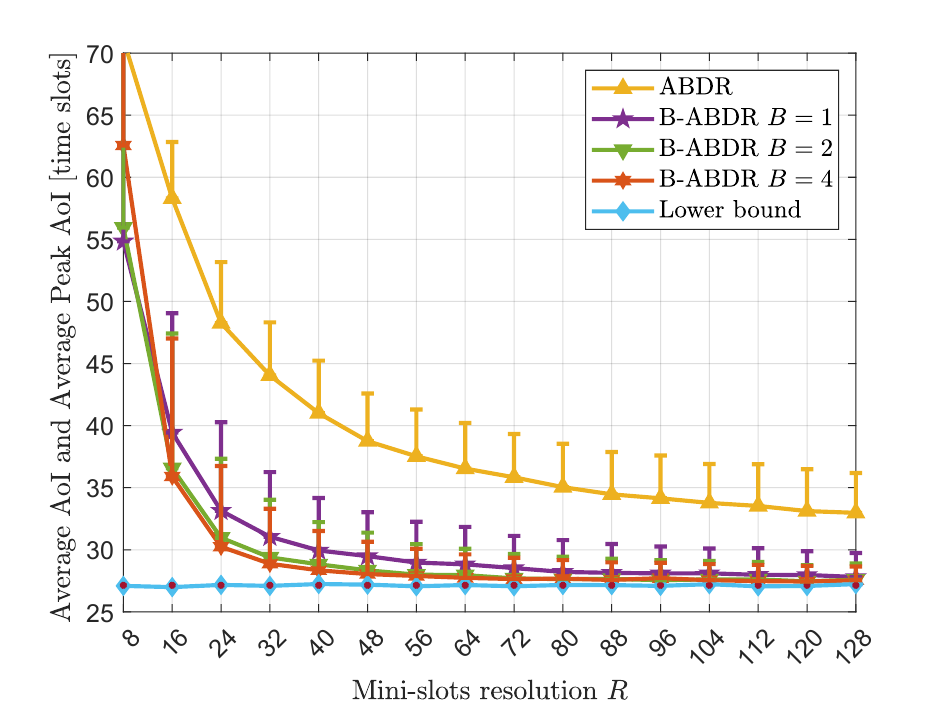}
    \caption{\changed{\gls{AAoI} and \gls{PAoI} as a function of the \gls{RTS} sub-slot resolution, $R$. Though the algorithms we propose achieve great performance with much less signaling, imperfections in the \gls{RTS} phase might affect them negatively, and proper configuration and resource allocation to the timers must be considered in practical settings.}}
    \label{fig:collisions_p2}
\end{figure}

\changed{
\subsection{B-/{ABDR} with Collisions in Phase-2}
We consider an implementation of \gls{B-ABDR} with possible collisions in phase-2. Here, the timer used by the relays to initiate the \gls{RTS} process is assumed to be synchronized, but time is discretized into mini-slots, with $R$ denoting the resolution of the timer, i.e., of how many mini-slots it is composed. When the earliest relays select the same mini-slot to start the \gls{RTS}, none can sense the other requests, and all such relays start a transmission, resulting in a potential collision. In Fig. \ref{fig:collisions_p2}, we present the \gls{AAoI} and \gls{PAoI} obtained with \gls{B-ABDR} as a function of $R$ for the default configuration presented in Table \ref{tab:list_of_symbols}. In a practical deployment, the time allocated for the RTS must be selected based on the timers available so that a satisfactory performance is achieved.}

\changed{
\subsection{Heterogeneous networks}\label{subsec:hetnet}
While we mainly address the problem of forwarding in phase-2 to decrease the \gls{AAoI} of the network in this paper, some characteristics of phase-1 can be investigated for a holistic protocol design that encompasses both phases. Our assumption that all \glspl{ED} are subject to the same erasure rate may not be representative of heterogeneous networks where \glspl{ED} do not have the same transmission power budget and/or are deployed in significantly different environments. To model these scenarios, we consider a network where each \gls{ED} is subject to a random value of $\varepsilon_1$ that follows a uniform distribution with parameters 0.05 and 0.5. We simulate 100 realizations of networks, each lasting for $10^5$ time-slots. In Fig. \ref{fig:hetnet}, we present a scatter plot of the \gls{AAoI} of all users at the end of the simulations, and a linear regression for each algorithm considered. Although the trends observed before are still present, the performance of users with a worse erasure rate is impacted. However, as B-ABDR is close to the lower bound, this is clearly a problem in phase-1, which inspires the extension of our proposal to also tackle phase-1. For instance, repetition strategies could be employed in the \glspl{ED} with the worse erasure rates, with the relays filtering out already received messages needing minimal feedbacks for the \glspl{ED}.

\begin{figure}
    \centering
    \includegraphics[width=\linewidth]{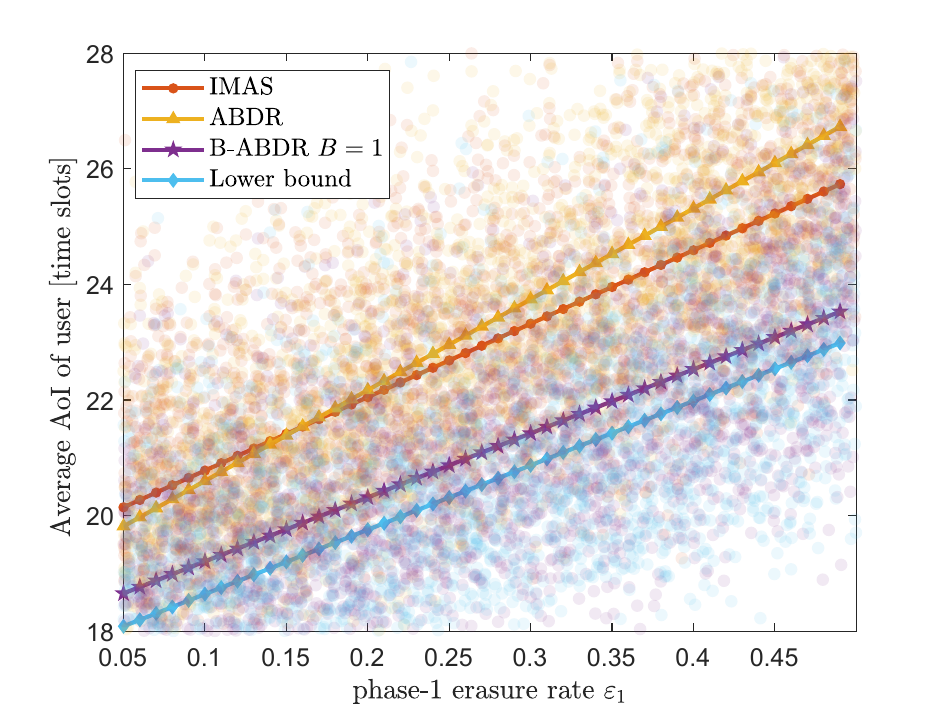}
    \caption{\changed{The AAoI versus the phase-1 erasure rate $\varepsilon_1$.}}
    \label{fig:hetnet}
\end{figure}
}

\section{Conclusions}\label{sec:conclusions}
In this work, we studied a two-hop multi-relay setup where relays use multiple frequency channels to forward information packets from \glspl{ED} to an \gls{AP}. We presented an analytical formulation for the minimal achievable \gls{AAoI} and \gls{PAoI} when the \glspl{ED} operate in ALOHA fashion, and validated the formulation with computer simulations. Furthermore, we extended two recent algorithms from the literature to the two-hop setup considered and proposed two new competing algorithms. The extended algorithms are \gls{MAM} and \gls{IMAS}, and they achieve performances close to the lower bound, although with considerable signaling between \gls{AP} and relays, as well as high computational complexity for the case of \gls{MAM}. On the other hand, our proposed algorithms, namely \gls{ABDR} and \gls{B-ABDR}, have performance comparable to the former two algorithms, while requiring less time for information exchange between the \gls{AP} and relays. Specifically, \gls{MAM} and \gls{IMAS} have similar \gls{AoI} performances in most configurations, outperforming \gls{ABDR}. The two have performances almost indistinguishable from the lower-bound when the phase-2 erasure rate is low, but with a more evident performance gap as this parameter increases, as is the case for \gls{ABDR}. On the other hand, \gls{B-ABDR}'s performance is the closest to the lower-bound in all configurations, including when phase-2 erasure rate is the highest, while requiring as much information exchange as \gls{ABDR}.

\changed{
For future works, as was briefly discussed in Subsection \ref{subsec:hetnet}, the system model can be modified to better represent realistic scenarios with asymmetrical connections between \glspl{ED} and relays, inspiring the design of protocols to promote fairness across \glspl{ED}.} \changed{The results in Subsection \ref{subsec:hetnet} also indicate that a limiting factor for the \gls{AoI} performance is the protocol adopted in phase-1. Therefore, other options, such as CSMA in phase-1, should be considered for a holistic design, though they also come with challenges, such as the hidden node problem.}
\changed{Moreover, the algorithms can be extended into the multi-hop scenario, which poses its own challenges. For instance, the communication between all nodes in each hop may not be guaranteed, and a broadcast style \gls{RTS} may be far from optimal. Also, each hop may incur in additional delay to a point where an \gls{ED} may already have new data to transmit before its previous packet even reaches the \gls{AP}.}

\begin{appendices}
\section{Proof of Theorem~\ref{th:lb}}\label{app:proof}
\begin{IEEEproof}  
An exact expression for the distribution of $\Delta$ is possible to be obtained by modeling the sum of the \gls{AoI} of \glspl{ED} as a single \gls{DTMC}, but modeling it is extremely complex. On the other hand, we can approximate $\Delta$ by considering that $\{\Delta_i\}$ are i.i.d. as
\begin{equation}\label{eq:C_func}
    \Delta(t) \sim \mathbf{\Pi}^{(N)}(\delta N),
\end{equation}
where
\begin{equation}\label{eq:conv}
     \mathbf{\Pi}^{(n)}(\delta) =  \underbrace{\mathbf{\pi}(\delta) * \mathbf{\pi}(\delta) * \dots * \mathbf{\pi}(\delta)}_{n\text{ times}}
\end{equation}
is the convolution of $\mathbf{\pi}(\delta)$ with itself $n$ times.
We recall that $P\triangleq{p\,Q}/({1-p\,Q})$. Then,
\begin{equation}\label{eq:Pi_n}
  \mathbf{\Pi}^{(n)}(\delta) = P^n\,\frac{(1-p\,Q)^{\delta}}{(n - 1)!}\frac{(\delta-1)!}{(\delta - n)!}u(\delta-n),
\end{equation}
where $u(\cdot)$ is the unit step function. We will proceed with proof by induction. First, we define $\mathbf{\Pi}^{(1)}(\delta) \triangleq  \pi(\delta)$, and redefine \eqref{eq:pdf_aoi_eds} as 
\begin{equation}
    \pi(\delta) = P{(1-p\,Q)^{\delta}}u(\delta-1),
\end{equation}
to make explicit that it is only valid for $\delta\geq1$. Then,
\begin{equation}
\begin{split}
    \mathbf{\Pi}^{(2)}(\delta) & = \Pi^{(1)}(\delta) * \pi(\delta) \\
                 & = P^2\sum_{m=-\infty}^\infty\frac{(1-p\,Q)^m}{(1-p\,Q)^{m-\delta}}u(m-1)u(\delta-m-1) \\
                 & = P^2(1-p\,Q)^\delta u(\delta-2)\sum_{m=1}^{\delta-1} 1 \\
                 & = P^2(1-p\,Q)^\delta(\delta-1)u(\delta-2),
\end{split}
\end{equation}
which is equivalent to \eqref{eq:Pi_n} for $n=2$. Now, assuming \eqref{eq:Pi_n} to be true, we have,
\begin{equation}
\begin{split}
    \mathbf{\Pi}^{(n+1)}(\delta) & = \Pi^{(n)}(\delta) * \pi(\delta) \\
                            & = \frac{P^{n+1}}{(n-1)!} \\
                            &\sum_{m=-\infty}^\infty\frac{(1-pQ)^m}{(1-pQ)^{m-\delta}}\frac{m! u(m-n)u(\delta-m-1)}{m(m-n)!}\\
                            & = \frac{P^{n+1}}{(n-1)!}(1-pQ)^\delta u(\delta-n-1) \sum_{m=n}^{\delta-1}\frac{(m-1)!}{(m-n)!} \\
                            & = \frac{P^{n+1}}{(n-1)!}(1-pQ)^\delta u(\delta-n-1) \frac{(\delta-1)!(\delta-n)}{n(\delta-n)!} \\
                            & = P^{n+1}\frac{(1-pQ)^\delta}{n!} \frac{(\delta-1)!}{(\delta-n-1)!} u(\delta-n-1). \\
\end{split}
\end{equation}
The same result can be obtained with \eqref{eq:Pi_n} by setting $n=n+1$, which proves that the equation is valid. Then, \eqref{eq:theorem} is obtained by scaling \eqref{eq:Pi_n} as \eqref{eq:C_func}, and changing the factorial operation to the Gamma function as now the domain is continuous. This concludes the proof.
\end{IEEEproof}  

\end{appendices}

\bibliographystyle{IEEEtran}
\bibliography{main}

\end{document}